\documentclass{aa}
\usepackage{natbib}

\begin{document}

\title{On the unpulsed radio emission from J0737-3039}

\author{R.~Turolla\inst{1} \and A.~Treves\inst{2}}

\titlerunning{Unpulsed radio emission from J0737-3039}
\authorrunning{Turolla and Treves}

\offprints{R.~Turolla, \email{turolla@pd.infn.it}}

\institute{
Dipartimento di Fisica, Universit\`a di Padova, Via Marzolo 8, I-35131
Padova, Italy\\
\and
Dipartimento di Fisica e Matematica, Universit\`a dell'Insubria,
Via Valleggio 11, I-22100 Como, Italy}

\abstract{
The double pulsar system J0737-3039 appears associated with a
continuous radio emission, nearly three times stronger than that
of the two pulsars together. If such an emission comes from a
transparent cloud its spatial extent ($\ga 10^{13}\, {\rm
cm}$) should be substantially larger than the orbital separation.
Assuming homogeneity and equipartition, the cloud magnetic field
is $\sim 0.03\, {\rm G}$ and the electron characteristic energy
$\sim 60\, {\rm MeV}$. This is consistent with supposing that
relativistic electrons produced in the shock formed by the
interaction of the more luminous pulsar wind with the
magnetosphere of the companion flow away filling a larger volume.
Alternatively, the unpulsed emission may directly come from the
bow shock if some kind of coherent mechanism is at work. Possible
observational signatures that can discriminate between the two
pictures are shortly discussed.
\keywords{Pulsars: individual (PSR J0737-3039A/B) -- Radiation
mechanisms: non-thermal -- Stars: neutron}
}

\maketitle


\section{Introduction}

The double pulsar system J0737-3039 (\citealt{burgay03};
\citealt{lyne04}) is unique in many respects. Out of the eight
known double neutron star (NS) systems, it is the only one where
both neutron stars (A and B) manifest themselves as radio
pulsars. The pulsars periods are $P_\mathrm{A}=0.02$\, s,
$P_\mathrm{B}=2.77$\,
s with period derivatives $\dot P_\mathrm{A}=1.74\times 10^{-18}\, {\rm
ss}^{-1}$, $\dot P_\mathrm{B}=0.88\times 10^{-15}\, {\rm ss}^{-1}$. The
orbital period $P_{\mathrm {orb}}=0.10$\, d is the shortest of the eight, and
makes the system an ideal laboratory to observe general
relativistic effects. The simultaneous, unprecedented measurement
of several Post-Newtonian parameters allows to place
stringent constraints on the NS masses ($M_\mathrm{A}\simeq 1.34\, M_{\sun}$,
$ M_\mathrm{B}\simeq 1.25 \, M_{\sun}$) even if the system has been monitored
for less than a year. The orbital separation of the two stars is
typically $d\sim 9\times 10^{10}$\, cm and the estimated distance
of J0737-3039 is $D\sim 600$\, pc (\citealt{lyne04}). Very
recently, a faint X-ray source ($L_\mathrm{X}\approx 2\times 10^{30}\, {\rm
erg\, s}^{-1}$) has been observed with {\em Chandra} at the position
of J0737-3039 (\citealt {mcl04}).

The total flux detected at 1390 MHz from J0737-3039 is $\sim 7$\,
mJy. The time-averaged pulsed flux from the two pulsars is $\sim
1.8$\, mJy indicating that the largest part of the system radio
emission is unpulsed (\citealt{lyne04}). This is the first time
that a continuous radio-source appears associated to an old pulsar
(with the possible exception of nearly aligned rotators, see e.g.
\citealt{hank93}). As suggested by \cite{lyne04}, the continuous radio
emission might be associated with the interaction of the two pulsar winds,
a situation realized in J0737-3039 alone. Assuming isotropic emission, an
unpulsed flux of $F_\mathrm{C}\sim 5$\, mJy at 1390 MHz corresponds to a
luminosity
$L_\mathrm{C}\sim 3\times 10^{27}\, {\rm erg\, s}^{-1}$ in the same band at
the
given distance of $0.6$\, kpc.

In this letter we investigate in more detail the nature of the
continuous radio emission from J0737-3039. Two possible scenarios
to account for the observed unpulsed flux are presented in
\S\ref{models} and observational signatures that can discriminate
between them  briefly discussed in \S\ref{discussion}.

\section{Continuous radio emission models}\label{models}

Because of the large difference in the spin-down luminosity of the
two pulsars ($\dot E_\mathrm{A}\sim 5.8\times 10^{33}\, {\rm erg\, s}^{-1}$,
$\dot E_\mathrm{B}\sim 1.6\times 10^{30}\, {\rm erg\, s}^{-1}$), the
relativistic wind of A penetrates deep into the magnetosphere of
B. A bow shock should be produced where the energy density
associated with the wind of A equals that of the magnetic field of
B. A simple calculation assuming a dipole field for pulsar B with
surface strength $B_\mathrm{B}\sim 1.6\times 10^{12}\, {\rm G}$ shows that
this happens at a distance $r_\mathrm{s}\sim 6\times 10^9 \, {\rm cm}$ from
B. Since B's light cylinder radius is $r_{\mathrm{c,B}}\sim 1.3\times
10^{10}\, {\rm cm}$, the shock is well within $r_{\mathrm{c,B}}$
\citep{lyne04}.

The magnetic field of pulsar B at the shock is
$B_{\mathrm{s}}\approx 7\, {\rm G}$. Assuming that the linear
dimension of the shock is $\approx r_\mathrm{s}$, and its width is
$\eta r_\mathrm{s}$ with $\eta \ll 1$, it is easy to show that the
synchrotron depth is larger than unity at radio frequencies if the
density of relativistic electrons produced in the shock itself
exceeds $n_\mathrm{e} \approx 10^{2}\, {\rm cm}^{-3}$ (see
\S\ref{thick}). As noted by \cite{kaspi04}, such an opaque plasma
sheath is probably responsible for the eclipses of pulsar A at
certain phases (see also \citealt{demorest04}).

Despite the magnetosheath is definitely the site of particle
acceleration and hence of synchrotron emission, the fact that it
is thick to radio photons implies that the released power is {\em
prima facie} orders of magnitude below $L_\mathrm{C}$. This brings in the
question
of how and where the continuous radio emission is produced. In the
following we discuss two possible scenarios for explaining the
continuous flux. The first is based on the assumption that the
radio-source is transparent (or quasi-transparent) to radiation at
1390 MHz. The second considers the possibility that the radio
emission comes from the bow shock but it is coherent.

\subsection{The transparent scenario}\label{thin}

First we consider a homogeneous spherical cloud and derive the
basic physical parameters in the hypothesis that the cloud is
transparent at radio frequencies ($\nu\approx 1400$\, MHz), and
that the magnetic and relativistic electron energy densities are
in equipartition. We further assume that there is a characteristic
electron Lorentz factor $\gamma$, and that the energy spectrum
around $\gamma$ has a slope $p=2$. Let $N_\mathrm{e}$ denote the total
electron number, $R$ the cloud radius, and $B$ the magnetic field
strength. From the condition that the typical frequency is of the
order of the synchrotron frequency $\nu_\mathrm{c}\sim 4.2\times 10^6
B\gamma^2$\, Hz, it follows

\begin{equation}\label{gamma}
\gamma\sim 18B^{-1/2}\, .
\end{equation}
Equipartition between magnetic and relativistic particles energy
density is
\begin{equation}\label{equipart}
\frac{B^2}{8\pi}\sim\left(\frac{3N_\mathrm{e}}{4\pi R^3}\right)\gamma
m_\mathrm{e}c^2
\end{equation}
from which it follows
\begin{equation}\label{B_eq}
B\sim 0.02 N_\mathrm{e}^{2/5}R^{-6/5}\, {\rm G}
\,.
\end{equation}
The total synchrotron power emitted by a single electron is

\begin{equation}\label{p_e}
P_\mathrm{e}\sim  10^{-15}\gamma^2B^2\sim 4\times 10^{-13}B\,
{\rm erg\, s}^{-1}\, .
\end{equation}
By equating the total luminosity $N_\mathrm{e}P_\mathrm{e}$ to that of the
continuous
radio emission,
$L_\mathrm{C}\sim 3\times 10^{27}\, {\rm erg\, s}^{-1}$, we get

\begin{equation}\label{N_e}
N_\mathrm{e}\sim 5\times 10^{29}R^{6/7} \,.
\end{equation}
The monochromatic synchrotron absorption coefficient for electrons
with a power-law energy distribution of index $p$ is (e.g. \citealt{rl})

\begin{equation}\label{alpha1}
\alpha_\nu\sim 2\times 10^{17}\frac{3N_\mathrm{e}}{4\pi R^3}
B^{(p+2)/2} \nu^{-(p+4)/2}\, {\rm cm}^{-1} \,.
\end{equation}
Using expressions (\ref{B_eq}) and (\ref{N_e}), taking
$\nu=1400$\, MHz and $p=2$, eq. (\ref{alpha1}) becomes

\begin{equation}\label{alpha2}
\alpha_{1.4\, {\rm GHz}}\sim 3\times 10^{39}R^{-27/7}\,
{\rm cm}^{-1}\, .
\end{equation}
The condition for marginal transparency is $\tau_\nu\sim\alpha_\nu R\sim 1$
from which a minimum cloud size can be derived

\begin{equation}
R\approx 5\times 10^{13}{\rm cm} \,. \nonumber
\end{equation}
Correspondingly, the other parameters take the values
\begin{eqnarray}
& N_\mathrm{e} \approx 3\times 10^{41}, & B \sim 0.03\, {\rm G},\cr\nonumber
& &\cr
& n_\mathrm{e} \approx 0.6\, {\rm cm}^{-3}, & \gamma \approx 100\,.
\nonumber
\end{eqnarray}
A homogeneous cloud model is obviously far from giving a realistic
description of the source, and the parameters given above should
be taken just as terms of reference.

If we assume that the magnetic field decays like $1/r^3$ inside
$r_\mathrm{c}$ and as $1/r$ in the radiation zone, it is noticeable that
at $r\approx 10^{13}\, {\rm cm}$ the field expected from pulsar A
is $B\sim 7\times 10^{-2}\, {\rm G}$, comparable to the
equipartition value employed above. It is then reasonable that
the field generated by pulsar A itself accounts for the synchrotron
emission.

Since a continuous radio flux does not appear in isolated pulsars,
but it is a unique characteristic of this system, the relativistic
electrons responsible for the synchrotron emission are most
probably produced at the shock. In this respect we note that the
unpulsed radio luminosity is only a small fraction of the wind
luminosity of A intercepted by the bow shock, $L_\mathrm{S}\approx
[r_\mathrm{s}/(d-r_\mathrm{s})]^2\dot E_\mathrm{A}\approx 3\times
10^{31}\, {\rm erg\, s}^{-1}$, so there will be plenty of energy
to accelerate the electrons. If we refer again to the parameters
of the homogeneous model, supposing an isotropic distribution of
pitch angles and a typical particle energy $\sim \gamma
m_\mathrm{e} c^2\sim 60\, {\rm MeV}$, the timescale for
synchrotron losses is $\tau_\mathrm{synch}\approx m_\mathrm{e}
c^2\gamma /P_\mathrm{e}\approx 10^{10}\, {\rm s}$. This means that
electrons accelerated at the shock continuously flow to larger
distances. An effective drift distance of $\sim 10^{13}\, {\rm
cm}$ is consistent with a mean free path (taken to coincide with
the Larmor radius) of $\sim 10^7\, {\rm cm}$. \cite{zl04}
have recently estimated the total rate at which particles are
deposited in the bow shock by the wind of A to be $\dot N\approx
10^{35}\, {\rm s}^{-1}$ under typical conditions; the rate at
which particles leak from the shock into the magnetosphere of B is
ten times smaller and is neglected here. The total number of
particles injected during a synchrotron timescale is $\approx \dot
N\tau_\mathrm{synch}\approx 10^{45}\gg N_\mathrm{e}$. This should
ensure that enough relativistic particles are present even if the
pair multiplicity of A is below $\sim 10^6$ (see again
\citealt{zl04}) or only a fraction of the particles are actually
accelerated in the bow shock.

\subsection{The coherent scenario}\label{thick}

Since we now refer to particles populating the magnetosheath, we
take $B\sim 7\, {\rm G}$, the field strength of pulsar B at the
shock position $r_\mathrm{s}$. Now the typical volume of the
emitting region is $\eta r_\mathrm{s}^3$, where $\eta
r_\mathrm{s}$ is the shock width. The limiting frequency for
synchrotron self-absorption can be computed as a function of the
electron number $N_\mathrm{e}$ from the condition $\tau_\nu\sim
\alpha_\nu\eta r_\mathrm{s}\sim 1$. Using for the absorption
coefficient the expression given in (\ref{alpha1}), with $4\pi
R^3/3$ replaced by $\eta r_\mathrm{s}^3$ and again $p=2$, we get
\begin{equation}\label{nuthick}
\nu_{\mathrm{th}}\sim 5\times 10^{5}r_\mathrm{s}^{-2/3}
B^{2/3}N_\mathrm{e}^{1/3}\sim 0.6 N_\mathrm{e}^{1/3}\, {\rm Hz} \,.
\end{equation}
We note that, having expressed $\nu_{\mathrm{th}}$ in terms of the electron
number $N_\mathrm{e}$, eq. (\ref{nuthick}) is independent of $\eta$.

The spectrum of a self-absorbed synchrotron source is peaked
around $\nu_{\mathrm{th}}$ (e.g. \citealt{rl}), so, as a first
approximation, we can assume that the total luminosity is given by
\begin{equation}\label{lthick}
L\approx \nu_{\mathrm{th}}L_{\nu_{\mathrm{th}}}\, ,
\end{equation}
where $L_{\nu_{\mathrm{th}}}$ is the monochromatic luminosity at
$\nu_{\mathrm{th}}$.
The latter can be found multiplying the synchrotron emission coefficient
times the volume of the emitting region
\begin{equation}\label{lnu}
L_{\nu}\sim \eta r_\mathrm{s}^3 P_\nu\sim 5\times 10^{-19}N_\mathrm{e}
B^{3/2} \nu^{-1/2}\, {\rm erg\, s}^{-1}\, {\rm Hz}^{-1}\, .
\end{equation}
Inserting eqs. (\ref{nuthick}) and (\ref{lnu}) into eq. (\ref{lthick}), we
get
\begin{equation}\label{lthick1}
L\approx 2\times 10^{-19} B^{11/6}N_\mathrm{e}^{7/6}\sim 7\times
10^{-18}N_\mathrm{e}^{7/6}\, {\rm erg\, s}^{-1}\, .
\end{equation}

From the energetics it follows that the total luminosity produced
by the magnetosheath is $L\la L_\mathrm{S}\approx 3\times 10^{31}\,
{\rm erg\, s}^{-1}$. The total particle number required to produce
such a power is $N_\mathrm{e}\approx 5\times 10^{41}$ (eq.
[\ref{lthick1}]). The emitted flux is peaked at $\approx
\nu_{\mathrm{th}}\approx 5\times 10^{13}\, {\rm Hz}$, four orders of
magnitude above the radio frequencies at which the continuous
emission has been detected. The luminosity in the GHz range will
be $\approx L_\mathrm{S}(10^9/\nu_{\mathrm{th}})^{5/2}\approx 6\times
10^{19}\, {\rm erg\, s}^{-1}$, many orders of magnitude below $L_\mathrm{C}$.
For the limiting
frequency given by eq. (\ref{nuthick}) to fall into the radio
domain, $\nu_{th}\approx 10^9\, {\rm Hz}$, it has to be
$N_e\approx 10^{28}$ which gives a total luminosity $L\approx 4\times
10^{15}\, {\rm erg\, s}^{-1}$.
This shows that synchroton emission from the  optically thick layer
can not be responsible for the observed unpulsed luminosity.

The brightness temperature associated with the unpulsed flux is

\begin{eqnarray}\label{tb}
T_\mathrm{b} &\sim&  10^{34} \left(\frac{F_\mathrm{C}}{1\, \mathrm{Jy}}\right)
\left(\frac{D}{1\, \mathrm{kpc}}\right)^2 \left(\frac{\nu}{1\, \mathrm{GHz}}
\right)^{-2}R^{-2}\, \mathrm{K}\, \cr
\noalign{\smallskip}
& \approx & 2\times 10^{15}\eta^{-2}\, \mathrm{K}
\end{eqnarray}
where $R\approx \eta r_\mathrm{s}$ is the size of the emitting
region. Would the unpulsed emission came from a population of
relativistic electrons, particles with a typical Lorentz factor
$\gamma\sim k_\mathrm{B}T_\mathrm{b}/m_ \mathrm{e}c^2 \ga 10^5$
should be present. These electrons should be confined in outermost
layers of the magnetosheath and their number is not to exceed
$N_\mathrm{e} \approx 10^{28}$ to avoid severe synchrotron
absorption. If the main radiative losses are through curvature
radiation, the luminosity in the radio range is given by

\begin{equation}\label{curvrad}
L_\mathrm{C}\approx L_\mathrm{S}\left(\frac{\nu_\mathrm{1\, GHz}}
{\nu_\mathrm{crit}}\right)^{4/3}
\end{equation}
where $\nu_\mathrm{crit}=c\gamma^3/2\pi\rho$ and $\rho$ is the curvature
radius of the trajectory (see e.g. \citealt{pare70}). By setting
$\rho\approx\eta r_\mathrm{s}$, it is  $\nu_\mathrm{crit}\ga 1\times 10^{7}
\, \mathrm{GHz}$ from which it follows from eq. (\ref{curvrad}) that
$L_\mathrm{C}\la 10^{-9}L_\mathrm{S}\sim 10^{22}\, \mathrm{erg\, s}^{-1}$.

It seems therefore that the only possibility left to explain the
unpulsed emission in terms of emission from the magnetosheath is
to invoke a coherent mechanism. Coherent emission has been
extensively investigated in connection with radio pulsars and
comes into three types: maser, reactive and ``coherent'', or
emission by bunches (e.g. \citealt{rusu75}; \citealt{mel78};
\citealt{zhang99} and references therein). In the latter case
particles in a bunch with spatial scale smaller than a wavelength
radiate in phase and the total power emitted by a single particle
is enhanced by a factor $N$, the number of particles in a bunch.
Despite applications of the bunching mechanism to pulsars' radio
emission have been criticized in the past (e.g \citealt{mel78}),
recent investigations have shown that a free-electron laser (FEL)
can indeed be operating high up in the pulsar magnetosphere, where
the background magnetic field is $\la 100\, \mathrm{G}$
(\citealt{fuku04} and references therein). Although the proposed
scenario for the FEL is quite different from the present one, one
may speculate that the same basic mechanism is at work in the two
cases. A further possibility to achieve coherent emission, as
recently suggested by \cite{zl04}, is the two-stream instability
which may develop between the downstream wind of pulsar A and the
upstream wind of B.

\section{Discussion}\label{discussion}

While the transparent scenario is rather conventional, the second
scenario postulates the presence of electron bunching and of some
coherence in the radio emission. The only argument we can quote
for the latter situation is that coherence needs to be
required for explaining pulsars radio emission, as it follows from
elementary considerations on the radio brightness temperature.
The geometric and physical conditions where the pulsed and continuous
emission described in \S\ref{thick} arise may be somehow similar,
although we are aware that the analogy is only tentative
and that electron bunching is just one of the possible mechanisms.

The two pictures proposed above lead to rather different
observational expectations. The transparent radio source has an
apparent diameter $\ga 1\, {\rm mas}$ at the putative distance
of $\sim 600\, {\rm pc}$ and could be, in principle, resolved by radio
telescopes like VLBA. In the coherent model variability on timescales
$\sim 0.1\, {\rm s}$ is expected together with some degree of orbital
modulation, produced by the change in the bow shock aspect ratio
with phase. No modulation at all should be present in the
transparent picture. It is worth noting that the transparent cloud
can not be responsible for the X-ray emission detected by {\em
Chandra} \citep{mcl04}. In fact, for a typical energy of primary
photons $\approx h\nu_\mathrm{c}\approx 6\times 10^{-6}\, {\rm eV}$,
inverse Compton on electrons with $\gamma\approx 130$ will produce
at most IR radiation. On the other hand, the X-ray luminosity could be
produced at the shock or by pulsar A itself.

\begin{acknowledgements}
We are grateful to A. Possenti for several useful discussions. Work
partially supported by the Italian Ministry
for Education, University and Research (MIUR) under grant
PRIN-2002027145.
\end{acknowledgements}

\end{document}